# Is There a Fourth Law for Non-Ergodic Systems
# That Do Work to Construct Their Expanding Phase Space?

Stuart Kauffman
University of Pennsylvania, Professor Emeritus, Department of Biophysics and Biochemistry

**Email:** stukauffman@gmail.com

**Abstract**

Substantial grounds exist to doubt the universal validity of the Newtonian Paradigm that requires a pre-stated, fixed phase space. Therefore, the Second Law of Thermodynamics, stated only for fixed phase spaces, is also in doubt. The validity of the Newtonian Paradigm may stop at the onset of evolving life. Living cells and organisms are Kantian Wholes that achieve constraint closure, so do thermodynamic work to construct themselves. Evolution constructs an ever-expanding phase space. Thus, we can ask the free energy cost per added degree of freedom. That cost is roughly linear or sublinear in the mass constructed. However, the resulting expansion of the phase space is exponential. Thus, the evolving biosphere does thermodynamic work to construct itself into an ever-smaller sub-domain of its ever-expanding phase space at ever less free energy cost per added degree of freedom. Entropy really does decrease.

A testable implication of this, termed here the Fourth Law of Thermodynamics, is that at constant energy input, the biosphere will construct itself into an ever more localized subregion of its ever-expanding phase space. This is confirmed. The energy input from the sun has been roughly constant for the 4 billion years since life started to evolve. The localization of our current biosphere in its protein phase space is at least $10^{-2540}$. The localization of our biosphere with respect to molecules of CHNOPS comprised of up to 350,000 atoms is extremely high. Entropy has decreased. The universality of the Second Law fails.

*"If someone points out to you that your pet theory of the universe is in disagreement with Maxwell's equations – then so much the worse for Maxwell's equations. If it is found to be contradicted by observation – well, these experimentalists do bungle things sometimes. But if your theory is found to be against the second law of thermodynamics, I can give you no hope; there is nothing for it but to collapse in deepest humiliation."*

— Sir Arthur Eddington

Scientific Folk Wisdom:

"If a famous old scientist tells you something is impossible, he is invariably wrong."

With some wry amusement, my aim in this article is to demonstrate that wonderful Eddington is, in fact, wrong. Entropy can decrease in biospheres that can do thermodynamic work to expand their phase space. Sir Arthur Eddington is also correct. The Second Law of Thermodynamics is the most well- established theory in classical physics. Disorder—entropy—tends to increase. Given the time reversibility of the fundamental laws of classical and quantum physics, the Second Law of Thermodynamics is widely held to be the Arrow of Time.

But must it be true? The conceptual foundations of the Second Law are two claims: *i)* The Newtonian Paradigm: the system is in a pre-stated and fixed phase space (1); and *ii)* The Ergodic Hypothesis: the system spends equal time in equal volumes of its phase space (2).

Both claims are central to classical and quantum physics. Here is the Newtonian Paradigm (1). First, state the relevant variables. For Newton these are position and momentum. Next, state the laws of motion in differential form coupling the relevant variables. For example, for Newton there are his three Laws of Motion and Universal Gravitation. Third, define the boundary conditions of the system. These boundary conditions thereby define the pre-stated and fixed "phase space" of all possible combinations of the values of the relevant variables. Fourth, state the initial conditions. Finally, integrate the equations of motion to obtain the entailed trajectory of the system in its fixed phase space. For classical physics this is a trajectory. For quantum physics, on most interpretations, it is an entailed trajectory of a probability distribution.

Here is the Ergodic Hypothesis (2). The system spends equal times in equal volumes of its *fixed* phase space. The Ergodic hypothesis abandons integrating Newton's equations of motion for the $N$ particles in the box. Given $N$ particles, $6N$ numbers specify the positions and momenta of all $N$ particles in the *fixed* 6N dimensional phase space. Given the Ergodic Hypothesis, the system will spend equal times in equal volumes of the fixed phase space so will spend more time in macrostates with more tiny $6N$ dimensional microstates. Hence disorder will tend to increase, and entropy, the logarithm of the number of microstates in a macrostate, will tend to increase. This is the Second Law and the proposed Arrow of time.

Under the Second Law, the *localization* of the system in its fixed phase space *tends to decrease*. It is of fundamental importance that the Second Law depend upon the Newtonian Paradigm of a fixed phase space.

**Is the Newtonian Paradigm Universal?**

Does the universality of the Newtonian Paradigm hold? There are increasing grounds for doubt. Smolin (1) points out the no mathematical object can entail the evolution of the universe because the entailing relation is timeless, but time and Now are real. A timeless mathematical object has a fixed phase space. If time and Now are real, there can be no fixed phase space. Therefore, the Newtonian Paradigm cannot be universal. Devereaux and colleagues (3) have recently shown that no modeler inside the universe can have a complete model of the universe. If correct, this rules out a pre-stated and fixed phase space, hence rules out a Final Theory that entails the evolution of the universe: The Newtonian Paradigm is thus not universal.

Kauffman and Roli have shown that it is not possible to use any mathematics based on set theory to deduce the ongoing evolution of ever-new adaptations in the evolution of life (4). The evolving phase space of the biosphere includes these ever-new adaptations. Thus, the ever-changing phase space cannot even be deduced. There is no fixed phase space. The corollary is that there can be no final theory in a universe containing at least one evolving biosphere (5). The universality of the Newtonian Paradigm again fails. The clear implication of Kauffman and Roli is that the phase space of evolving life is not fixed (4,5). The vast increase in the abundance and diversity of life in the past 4 billion years hints, but does not yet prove, that the phase space of the evolving biosphere has increased.

**The Universe is non-Ergodic**

It has become clear for some time that the Universe is non-ergodic on time scales vastly longer than the lifetime of the universe (6,7). Consider encoded proteins found in prokaryotic and eukaryotic cells. A typical protein has about 350 amino acids arranged in peptide bonds. Then consider a shorter protein with 200 amino acids. How many proteins of length 200 are possible? Assuming 20 different encoded amino acids, there are $20^{200} = 10^{260}$ possible proteins length 200 amino acids. Could all of these have been synthesized in the lifetime of the universe?

The shortest time scale is the Planck time scale of $10^{-43}$ seconds. The universe is $10^{17}$ seconds old. There are an estimated $10^{80}$ particles in the universe. If all these particles, ignoring space-like separation, were creating proteins length 200 on the Planck time scale, it would require the age of universe raised to the 37$^{th}$ power to make all these possible proteins just once (6,7). Therefore, at the scale of complex organic molecules such as proteins with 200 amino acids, the universe is vastly nonergodic. In fact, the universe is not ergodic above about 500 daltons (8).

**The Biosphere Has Vastly Expanded its Phase Space**

Another basis to doubt the universality of the Newtonian Paradigm comes from a recent analysis. General relativity and cosmology both assume this paradigm to be foundational. Cortês *et al.*, (9,10,11) undertook the analysis of the complexity of a biosphere compared to that of the abiotic universe, in order to assess the implications of the existence of life in the universe on the "price" for the initial state of the universe.

The price for the initial state is the Past Hypothesis: Given; *i*) the Newtonian Paradigm with its fixed and unchanging phase space; *ii*) the Second Law in which disordered complexity increases; and *iii*) the present complexity of the universe; then, *the initial state* of the universe must have been of correspondingly very low entropy. The present complexity of the *abiotic* universe is estimated to be an enormous $e^{10^{10^{124}}}$. The entropy of the initial state was then the reciprocal: $1/e^{10^{10^{124}}}$. Penrose points out how difficult is this price for the initial state and the Past Hypothesis to be paid (12).

The work of Cortês *et al.* (9, 10,11) estimating the complexity of our single biosphere up to the first encoded protein synthesis makes this price very much higher. The price is not a "mere" reciprocal of $e^{10^{124}}$ but for a single biosphere *the price is the reciprocal of* $10^{10^{237}}$, where *237 >> 124.* Because $10^{10^{237}}$ for our biosphere is vastly greater than $e^{10^{124}}$ for the entire abiotic universe, we must conclude that the phase space of the biosphere has, in fact, expanded enormously.

Here are our choices for a cosmology that includes biospheres, *i.e.*, a "biocosmology" (9, 10,11). First, we can choose to deny the results of Cortês *et al.* Second, we can accept the Cortês results and choose to maintain the universality of the Newtonian Paradigm and Second Law. We then preserve current cosmology. But in doing so, we al*so preserve a required price for the initial condition*, and we agree to pay the now vastly higher price. Third, the price of the initial state is conditioned on truth of the universality of the Newtonian Paradigm and the Second law. Thus, as our third choice, we can choose to abandon either or both of the universality of the Newtonian Paradigm and the universality of the Second law for the cosmological evolution of the universe. Because the Second Law depends upon the Newtonian Paradigm of a pre-stated and fixed phase space, giving up the universality of the Newtonian Paradigm calls the Second Law itself into question.

This article considers a candidate *Fourth Law of thermodynamics for non-ergodic systems such as evolving biospheres that can do thermodynamic work to construct their own expanding phase spaces*. In such an expanding phase space, perhaps astonishingly, order can increase, in flat contradiction to the Second Law. The system constructs itself into an ever-smaller region of its ever-expanding phase space.

Given the three choices above, I here take the third choice: Abandon the universality of Newtonian Paradigm with its fixed and pre-stated phase space. There is a "conservative" way to abandon the universality of the Newtonian Paradigm. Claim that: *with the onset of life, evolving biospheres create new possibilities that expand a phase space that is no longer fixed, (4,5,9,10,11). But prior to the onset of life, the Newtonian Paradigm holds, from the Big Bang onward with its fixed phase space.* In this case, the problem of the Past remains. The complexity of the current abiotic universe is estimated to be $e^{10^{124}}$. The initial state must pay the price. It is localized to the reciprocal of $e^{10^{124}}$. Penrose's dismay persists (12).

Remarkable independent grounds exist to support the "conservative" way to abandon the universality of the Newtonian Paradigm only at the onset of evolving life. Paul Davies, in 2004 published a paper entitled, "Emergent Properties and the Computational Properties of the Universe" (13). Davies argues that any physical law must be implemented within the resources of the universe. Given a maximum rate of elementary operations, $2E/\pi$, and that time starts at the Big Bang, he concludes that "an upper bound for the total number of bits information that have been processed by all the matter in the universe is … $10^{120}$. Expressed informally, the existence of an emergent law in a system of sufficient complexity that its behavior could not be described or predicted by processing $10^{120}$ bits of information will not come into conflict with any causal closure at the microlevel." Taking account of Dark Energy, Davies raises the limits to about $10^{122}$. Davies concludes that proteins longer than 60–90 amino acids, and nucleic acids longer than 200 nucleotides are open to emergent behavior not determined by any causal closure at the microlevel. Finally, Davis notes that many proteins are far longer than 90 amino acids and many genes are far longer than 200 nucleotides, so emergence is not ruled out (13).

Davies arguments are entirely consistent with the complexity of the abiotic universe found by Cortes *et al*.: $e^{10^{10^{124}}}$. Taken together, the arguments of Davies and of Cortes *et al*. support the claim that the *Newtonian Principle holds for the abiotic universe,* however the vastly greater complexity of the evolving biosphere, $10^{10^{10^{237}}}$, now becomes strongly positive evidence for emergence with the onset of life.

More, the huge excess of $10^{10^{10^{237}}}$ versus $e^{10^{10^{124}}}$ suggests that emergence in the evolving biosphere has been extremely important. The remainder of the article suggests some of the reasons for this.

**Thermodynamic Work Has Been Done to Expand the Evolving Biosphere's Phase Space**

Because the universe is non-ergodic above about 500 daltons, most complex things will never exist. Yet the human heart, 300 grams and able to pump blood, exists in the universe. How can this have become true (7,8)? To discuss this broad topic, I must explore ten issues:

1. I ask my physicist colleagues to consider the question above. How indeed have hearts come to exist in the non-ergodic universe? The basic answer is that life emerged and evolved. Hearts pump blood that sustains the life of organisms with hearts. Organisms with hearts have offspring that also having hearts. Those organisms whose hearts function better at sustaining the whole organism, have more offspring. Natural selection selects for improve hearts. Organisms with hearts evolve. Thus, hearts exist in the non-ergodic universe (7).

2. Organisms are Kantian Wholes. A Kantian whole has the property that the Parts exist for and by means of the Whole. The human reader of this article is a Kantian Whole. You exist for and by means of your parts: your heart, liver, kidneys and other organs and cells. They exist for and by means of you, the Kantian Whole (7).

3. The simplest example of a Kantian Whole is a *Collectively Autocatalytic Set.* Gonen Ashkenasy has a set of nine small peptides, 1, 2, ... 9. Each peptide binds and ligates two fragments of the next peptide into a second copy of the next peptide. Peptide 1 catalyzes by ligation a second copy of peptide 2. Peptide 2 catalyzes by ligation a second copy of peptide 3, and so on around a ring such that peptide 9 catalyzes by ligation a second copy of peptide 1. The system is collectively autocatalytic. No peptide catalyzes its own formation (13) The Kantian Whole is the entire set of nine peptides that constitute its parts (14, 7).

4. The existence of Kantian Wholes in the non-ergodic universe permits a non-circular definition the "*function*" of a Part in the Kantian Whole. *The function of a Part is that subset of its causal properties that sustains the Whole.* The function of peptide 1 is to catalyze the formation of a second copy of peptide 2. If peptide 1 jiggles water in the petri plate that is a side effect, not its function. The function of the heart is to pump blood, not jiggle fluid in the pericardial sac or make heart sounds. Functions are real in the universe. The function of the heart is why it exists in the universe (7).

5. Living cells and organisms achieve the property of *constraint closure* (15). This property lifts life, based on physics, above physics in entirely unexpected ways. To wit, work is force acting though a distance. Atkins points out that, "*Work is the constrained release of energy into a few degrees of freedom*" (16). Consider a cannon, cannon ball at the base of the cannon, and powder between the base of the cannon and the cannon ball. The cannon is the constraint and is also a boundary condition. When the power explodes the cannon constrains the release of energy in the expanding gas to expand only along the bore of the cannon. The expanding gas does thermodynamic work on the cannon ball which is shot from the cannon. Without constraints on non-equilibrium processes there can be no work.

6. A new question: At the Big Bang, there were no cannons. Where did the cannon come from? It took work to make the cannon! The Work-Constraint Cycle: No constraint, no work. No work, often, no constraint (17). To envision constraint closure, consider three non-equilibrium processes, **1**, **2**, and **3**. And consider three constraints, **A**, **B**, and **C**. **A** constrains the release of energy in process **1** that does work to construct **B**. **B** constrains the release of energy in process **2** that does work to construct **C**. And **C** constrains the release of energy in process **3** that does work to construct **A** (15). Constraint-closed systems do thermodynamic work to construct the very constraints on the release of energy into the few degrees of freedom that then constitutes the work that construct the very same constraints (15, 17). Constraint-closed systems literally do thermodynamic work to construct themselves.

7. Kantian Whole collectively autocatalytic sets achieve constraint closure (7). To realize this, consider Ashkenasy's nine-peptide collectively autocatalytic set (14). Each peptide binds the two fragments of the next peptide and acts as a ligase linking the two fragments together via a peptide bond into the larger peptide. Thermodynamic endergonic work is done in forming that peptide bond. The peptide acting as a ligase and binding the two fragments of the next peptide lowers the activation barrier to form the new peptide bond. Therefore, the peptide ligase acting as a catalyst is a constraining boundary condition that constrains the release of energy into the few degrees of freedom that constructs the next peptide. Each peptide is a constraint, (7, 15). Because each of the nine peptides acts as ligases for the formation of the next peptide around the ring of nine peptides, the entire system is a Kantian Whole that is also collectively autocatalytic and achieves constraint closure (7, 14). Thus, a collectively autocatalytic set achieves constraint closure and does work to construct itself as it reproduces itself (7,14).

8. Living cells as Kantian Wholes undergo heritable variation and natural selection. These create adaptations that are novel in the universe of possibilities. The evolving biosphere expands its phase space (4,5, 9,10,11,17). Both before and after the invention of encoded protein synthesis, life was capable of heritable variations that created ever-newer molecular, morphological, and behavioral adaptations. In turn these created ever-newer niches for ever-newer species. The species diversity of the biosphere has increased enormously despite small and large extinction events. Thus, the phase space of the biosphere has expanded. It is important to realize that selection acts at the level of the Kantian Whole, not its parts. And therefore, selection is downward causation. What survives is that which is fit in the current environment. This conclusion comes in opposition to *S*. Weinberg (18), "*it is not true that all the explanatory arrows point downward to particle physics.*"

9. The evolution of ever new adaptations that expand the phase space of the biosphere cannot be deduced. Adaptations are "opportunities" or "affordances" seized by heritable variation and natural selection. A given protein in a cell now used to bind a ligand can also come to be used to carry a tension load, or to transmit an electron. An engine block can be used as a paper weight and its corners are sharp and can be used to crack open coconuts. It is not possible to deduce from the use of an engine block as a paper weight that the same object can be used to crack open coconuts (4,5). Because the indefinite uses on any object, X, cannot be deduced and cannot be listed, no mathematics based on set theory can be used to deduce the evolution of the biosphere (4,5).

10. The implication of all the above is that evolution is a propagating construction that cannot be deduced, hence evolution is not an entailed deduction. No Law entails the evolution of the biosphere whose expanding (or contracting) phase space cannot be deduced (4,5). As a consequence, the evolving biosphere must lie entirely outside of the Newtonian Paradigm. In short, life is based on physics but beyond physics. There can be no "final theory" for the evolution of a universe having at least one evolving biosphere (4,5,17).

**A Statistical Mechanics of Non-Ergodic Systems with Expanding Phase Spaces**

A preliminary issue here is, how can we define the phase space for a non-ergodic system with an expanding phase space? The natural concept of the phase space of a non-ergodic system is the count all the possibilities that *might* have occurred at any time "*t*". Call this "$P_t$". In general, some *subset* of all the possibilities at time, *t*, $P_t$, will have *actually* occurred. Call this actualized subset "$A_t$". At any time *t*, the ratio, $R = P_t/A_t$ measures current non-ergodicity of the system at time *t*. Conversely, the reciprocal, $1/R$, measures the localization of the actual system in its total possible phase space at time *t*.

The temporal variation of $P_t$, $A_t$, $R$, and $1/R$, that is, the non-ergodicity and localization of the system as the total possible expands, (or contracts) is then *a candidate 4$^{th}$ law for non-ergodic system (9,10,11)*. Progress toward a candidate Fourth Law requires some new theory for how $P_t$, $A_t$, $R$ and $1/R$ vary with time. The TAP Process (9,10,11) is the first such mathematical theory.

**TAP, The Theory of the Adjacent Possible in the Universe.**

For the chemical evolution of the universe, of life in our biosphere, and of technology for the past 2,500,000 years, the process appears to be roughly described by a new equation, the TAP equation (9,10,11,19,20,21,22,23):

**Equation 1.** $$M_{t+1} = M_t + \sum_{i=1}^{M_t} \alpha^i \binom{M_t}{i} \quad , \quad 0 \leq \alpha \leq 1$$

In this equation, $M_t$ is the number of "things" in the system at time *t*. A "thing" could be a kind of molecule, a species in the biosphere, a tool in a technological system or even an idea. In this equation, choose an initial value of $\alpha$, for example $\alpha = 0.9$. Then use $\alpha$ raised to the $i^{th}$ power.

The behavior of this process is striking and predicts two unrelated distributions (11,19,20, 21). If the process starts with a rather small number of types of items, $M_0 = 10$, and $\alpha << 1.0$, and iterates, the number of types of items increases glacially for a long time then explodes upward in a "hockey-stick" pattern. In the continuous version the number of things reaches infinity at a finite time. The TAP process thus has a pole. The discrete version does not reach infinity but explodes very rapidly (11,20,21). The TAP process roughly fits the chemical evolution of the increasing number of different atoms and molecules in the universe over 13.8 billion years (22), the increasing number of different species over 4 billion years of biological evolution, and the increasing number of different technologies over the past 2.6 million years of hominid evolution (23,20).

We can also interpret $M_t$ to be the most complex thing produced at time $t$, for example having $M_t$ parts. Then the TAP process yields a glacial then hockey- stick explosive increase both in the number of items and their gradual then explosive differentiation into simple and more complex items (11,19,20,21). The TAP process roughly fits the increasing *complexity* of atoms and molecules in the universe over 13.8 billion years from no atoms to atoms, to ever larger molecules, (22). TAP roughly describes the evolution of the increasing complexity of living species over 4 billion years (23). TAP also roughly describes the cumulative evolution of complexity of tools in our technology since *Australopithecus* 2.8 billion years ago. *Australopithecus* had perhaps ten crude and similar stone tools. Our tools today range in complexity from needles to the International Space Station (19,20,21). The fact that the TAP process seems to fit two different distributions, suggests that it is capturing something quite fundamental about the long-term evolution of complexity in the universe.

**TAP is the First Candidate for a Fourth Law of Thermodynamics**

TAP allows us to compute the *Total Possible*, the *Actualized Possible* and the ratio of these, $R = T_p/A_p$ at any time $t$. The temporal evolution of $T_p$, $A_p$, and $R$ constitute the candidate $4^{th}$ law. Using TAP and setting $\alpha = 1.0$ yields the evolution of the *Total Possible* as a function of time. Setting $\alpha < 1.0$ and fixed, yields the time evolution of the *Actualized Possible*. TAP allows the calculation of $T_p$, $A_p$, and $R$ as a function of time.

**THE FOURTH LAW**

1. *The Fourth Law 4.1 states that $T_p$, $A_p$, and **R** can all increase with time* (9,10,11). The consequences are remarkable. The Fourth Law, 4.1, states that $T_p$, $A_p$, **R**, and 1/**R**, the localization of the system, *can* all increase with time. The very possibility, that $T_p$, $A_p$, and **R** *can* increase flatly contradicts the Second Law where *localization* can only tend to decrease with time. If, via 4.1, **R** can increase, then *localization, 1/**R**, can increase*.

2. *The Fourth Law 4.2 states that $T_p$, $A_p$, and **R** all tend to increase with time.* The Fourth Law, 4.2, is more remarkable. It says that $T_p$, $A_p$, and **R** do, in fact, *tend to increase* with time. Hence *localization, 1/**R**, also does tend to increase*.

We will show below that the temporal behavior of our evolving biosphere confirms both 4.1 and 4.2. Importantly, *when localization increases entropy deceases!*

**What is the Free Energy Cost per New Added Degree of Freedom?**

This is a new question. In the fixed phase space of the Second Law, no issue can come up with respect to the free energy cost per added degree of freedom. In the Fourth Law, the phase space does expand, thus it becomes relevant to ask the free energy cost per added degree of freedom. The free energy cost per added degree of freedom should be roughly linear in the mass of the new thing constructed.

Consider a biosphere where the longest peptide length is $N$. Let a slightly longer peptide length $N+1$ be created. One new peptide bond has been created so the expansion of the phase space has a free energy cost of creating a single new peptide bond. The free energy cost is, at most, roughly linear in $N$. Further, it is well established that allometric scaling of ¾ power with respect to mass exists across 27 orders of magnitude for many phyla. The number of heartbeats per lifetime is independent of the mass of the organisms. Here the free energy cost per added degree of freedom is independent of mass (24).

Consider a biosphere whose longest peptide in at time $t$ is $N$. The phase space is $20^N$. Let a new peptide length $N + 1$ arise. The new phase space is $20^{(N+1)}$. *The phase space has expanded* *exponentially. Importantly, this exponential expansion of the phase space required no more work and generated no more heat than that required to create the single longer peptide.*

*This result is a central conclusion: The phase space, R, increases exponentially. Therefore localization, 1/R, also increases exponentially, but the free energy cost per added degree of freedom is roughly constant. Therefore,* **entropy truly decreases**. *This flatly contradicts the Second Law.*

*The further implication is that adding the next new degree of freedom becomes ever cheaper as the phase space grows.*

**The Cost of Adding the Next Degree of Freedom is Less as Degrees of Freedom are Added**

Under the Fourth Law, this new phenomenon emerges. In fact, a major issue for more than a century has been the question of how the biosphere has managed to vastly increase in complexity in face of the Second Law. This is one of the major issues posed by Schrödinger in his famous book *What Is Life*. He asks if new laws of physics may be required (25). Even though no laws entail the evolution of the biosphere (4,5), it remains true that living cells and organisms, *via* constraint closure, do thermodynamic work to expand the phase space of the biosphere. If the Fourth Law, which *is* a candidate new law, is true, can we test it with respect to our biosphere?

The Fourth Law makes a powerful prediction: at a constant energy input a system such as our biosphere can do work to expand it phase space will construct itself into an ever more localized subregion of its ever-expanding phase space, in other words, **1/R** decreases. *More, order increases while entropy decreases.* Can this this true?

*The evolution of the biosphere confirms the new Fourth Law*. Life emerged some 4 billion years ago. The annual energy input from the sun is roughly constant. Evolving organisms achieve constraint closure and do ongoing thermodynamic work to construct themselves in a propagating process by which an ever-constructed evolving biosphere of new possibilities keeps emerging (4,5,7,17). The evolving biosphere does ongoing thermodynamic work to expand its own phase space.

The complexity of the biosphere is now enormous. But that complexity is vastly smaller than all the possible biospheres than might have occurred, **R**. This is readily assessed by considering the just the known molecular complexity of the current biosphere considering only proteins of 2000 or more amino acids (26). There are 20 raised to the 2000 power or $10^{2600}$ possible proteins of length 2000 amino acids. Thus, a realistic lower bound on the total possible phase space of our known biosphere today at the level of encoded proteins as legitimate physical degrees of freedom is $10^{2600}$. *On this independent and data-based estimate, the complexity of the total phase space of our biosphere in terms of proteins as bound states is $10^{2600}$.*

As Eigen suggested (27), perhaps $10^{60}$ actual proteins have been "tried" in 4 billion years. At constant solar energy input, the biosphere is vastly localized in the total molecular phase space, ***R***, it has constructed compared to the free energy cost to achieve this localization. Given Eigen's estimate, the localization of our actual protein biosphere with respect to its total possible protein phase space is greater or equal to $10^{60} / 10^{2600} = 1/ 10^{2540}$.

The claim of increasing localization treats proteins as the units of interest. Is this legitimate? Any given protein can be disassembled into its *N* atoms, then considered in a liter box of buffer and the standard 6*N* dimensional phase space. What is a protein? It is a specific macrostate that corresponds to the very small number of microstates consistent with the locations of its *N* atoms. The protein itself is highly ordered in its 6*N* dimensional phase space, largely a consequence of quantum mechanics and stable covalent bonds.

If we consider all the proteins in living organisms in the biosphere, and disassemble them into their total N atoms, this is a very large 6*N* dimensional phase space. The total macrostate of these N atoms assembled into all the proteins in the biosphere is localized in a tiny sub-volume of this very large 6*N* dimensional phase space.

Concentrating only on proteins ignores the diversity of possible complex organic molecules comprised of many atoms of carbon, hydrogen, nitrogen, oxygen, phosphorus, and sulfur, CHNOPS, in the actual biosphere. The largest encoded protein is mammals is Titin with 35,000 amino acids (28). Each amino acid has on average ten atoms of CHNOPS. Therefore, Titin has roughly 350,000 atoms of CHNOPS. *The phase space of all possible organic molecules* of the current biosphere consists of all possible molecules comprised of 1 atom of CHNOPS, 2 atoms of CHNOPS, 10 atoms, 1000 atoms, 100,000 atoms, 350,000 atoms of CHNOPS. The total number of all possible molecules up to 350,000 atoms of CHNOPS is an unfathomably enormous number, *X*. The molecular diversity of our biosphere may be $10^{15}$ to $10^{20}$. Then the *localization* of our biosphere in terms of organic molecules is *$10^{20}/X$ … which is an unfathomably small number*.

**The Fourth Law is Correct**

If the 4$^{th}$ Law is valid and biospheres are abundant in the universe, the overall course of the total entropy of the evolving complex universe may need to be re-examined. Is the total entropy increasing or decreasing after life starts in a universe whose biospheres can do thermodynamic work to expand their phase spaces? Cortês *et al*. (9,10,11), estimate the complexity of our biosphere to be $10^{10^{237}}$, vastly larger than the $e^{10^{124}}$ for the abiotic universe. And $10^{-2540}$, and $10^{20}/X$ are vastly localized. *If correct, the entropy of the universe really has decreased since the origin of life in the universe.*

The Fourth Law, 4.2 states that $T_p$, $A_p$, $R$ *tend* to increase over time. However, the TAP process proceeds *inexorably* upward (9,10). This is essentially unchanged if a first order loss term, $-\mu$, per item at each time step is included (9). An important limitation of the TAP process as a full version of a Fourth Law is that the variables created do not interact with one another. This is inadequate. In the evolving biosphere and global economy, species and goods create niches for one another. Species and goods go extinct, new species and goods emerge and flourish. Reasonable evidence suggests that both the evolving biosphere and econosphere are dynamically "critical", and they generate power law distribution of extinction events in the Phanerozoic record, and power law distributed "Schumpeterian gales of creative destruction" in the evolving economy, (29,30,31). These dynamics appear to be endogenous. Because of these interactions among the items of TAP, the process does *not proceed inexorably* upward, but *merely tends* to proceed upward.

**We Have Taken the Second Law to be the Cosmological Arrow of Time. Is it?**

If the reasoning and results noted above are correct, the Second Law is not Universal. It only applies to systems with fixed phase spaces. But this is untrue of the evolving biosphere. The reasoning and results of this article imply that the Newtonian Paradigm is not universal. With the loss of that universality, the Second Law is not universal. The implication is that the Second Law cannot be the Cosmological Arrow of time.

The further implication is that we need to rethink Cosmology itself. One approach to reconsidering Cosmology is to take non-locality as fundamental. The immediate implication is to flatly contract General Relativity with its locality (32). Starting with non-locality as fundamental naturally yields a quantum gravity with a quantum arrow of time that is independent of the Second Law. This is not the interest of the present paper but may offer an alternative to the Second Law as the cosmological arrow of time, (32). This alternative may be testable (32).

**Conclusions**

Substantial grounds exist to doubt the universal validity of the Newtonian Paradigm that requires a pre-stated, fixed phase space. Therefore, the Second Law, stated only for fixed phase spaces, is also in doubt. The universe is not ergodic on vastly long-time scales. Living cells and organisms are Kantian Wholes that achieve constraint closure and do thermodynamic work to construct themselves. Evolution constructs an ever-expanding phase space. Thus, we can ask the free energy cost per added degree of freedom. That cost is roughly linear or sublinear in the mass constructed. However, the resulting expansion of the phase space is exponential. Thus, the evolving biosphere does thermodynamic work to construct itself into an ever-smaller sub-domain of its ever-expanding phase space at ever less free energy cost per added degree of freedom. Entropy really does decrease. A testable implication of this, the Fourth Law, is that at constant energy input, the biosphere can construct itself into an ever more localized subregion of its expanding phase space. This is confirmed. The localization of our biosphere in its protein phase space is at least $10^{-2540}$. Entropy has decreased. The universality of the Second Law fails.

**Acknowledgements**

The author is grateful for very useful conversations with Andrew Liddle, Marina Cortês, Leon Glass, Andrea Roli, and Niles Lehman.

Right to Publish: The author declares he has the right to publish this material and no conflicting interests.

**References**

1. L. Smolin. *Time Reborn: From the Crisis in Physics to the Future of the Universe*. Houghton Mifflin Harcourt, 2013.
2. C. C. Moore, Ergodic theorem, ergodic theory, and statistical mechanics. *Proc. Natl. Acad. Sci.* 112, 1907-1911 (2015). https://doi.org/10.1073/pnas.1421798112
3. A. Devereaux, R. Koppl, S. Kauffman, A. Roli, An incompleteness result regarding within-system modeling (2022). Available at https://papers.ssrn.com/solKKi3/papers.cfm?abstract_id=3968077
4. S. Kauffman, A. Roli, The world is not a theorem. *Entropy* 23, 1467 (2021).
5. S. Kauffman, A. Roli, The third transition in science: Beyond Newton and quantum mechanics – A Statistical Mechanics of Emergence. *arXiv*(4)15271 [physics.soc-ph] (2021).
6. S. Kauffman. *Humanity in a Creative Universe*. Oxford University Press, 2016.
7. S. Kauffman. *A World Beyond Physics*. Oxford University Press, 2019.
8. S. Kauffman, D. Jelenfi, G. Vattay. Theory of chemical evolution of molecule compositions in the universe, in the Miller-Urey experiment and the mass distribution of interstellar and intergalactic molecules. *J. Theor. Biol.* 486, 110097 (2020).
9. M. Cortês, S. Kauffman, A. Liddle, L. Smolin. Biocosmology: towards the birth of a new science. Available at https://www.biocosmology.earth/. *arXiv*:2204.09378 **[astro-ph.CO]** (2022).

10. M. Cortês, M., S. Kauffman, S. A. Liddle, L. Smolin, Biocosmology: Biology from a cosmological perspective. Available at https://www.biocosmology.earth. *arXiv*:2204.09379 **[physics.hist-ph]** (2022).

11. M. Cortês, M., Kauffman, S., Liddle A. R., L. Smolin, Biocosmology and the Theory of the Adjacent Possible. Available at https://www.biocosmology.earth, and *arXiv* https://doi.org/10.48550/arXiv.2204.14115. (2022).

12. R. Penrose. *The Road to Reality: A Complete Guide to the Laws of the Universe.* Alfred Knopf, 2005.

13. P.C.W. Davies, Emergent biological principles and the computational properties of the universe, *Complexity* 10, (2004). https://doi.org/10.1002/cplx.20059

14. Z. Dadon, N. Wagner, G. Ashkenasy, The road to non-enzymatic networks, *Angewandte Chemie International Edition* 47 (33), 6128-6136 (2008).

15. M. Montévil, M. Mossio, Biological organisation as closure of constraints. *J. Theor. Biol.* 372, 179-191 (2015). https://doi.org/10.1016/j.jtbi.2015.02.029

16. P.W. Atkins. *The Second Law*. Scientific American Library, 1994.

17. S. Kauffman, Answering Schrödinger's "What is life?" *Entropy* 22, 815 (2020).

18. S. Weinberg. *Dreams of a Final Theory*. Knopf Doubleday Publishing Group, 1993.

19. R. Koppl, A. Deveraux, S. Valverde, R. Solé, S. Kauffman, J. Herriot, The evolution of technology (2021). https://papers.ssrn.com/sol3/papers.cfm?abstract_id=3856338.

20. R. Koppl, A. Deveraux, S. Valverde, R. Solé, S. Kauffman and J. Herriot. Explaining technology (2021). https://papers.ssrn.com/sol3/papers.cfm?abstract_id=3856338.

21. M. Steel, W. Hordijk, S. Kauffman. Dynamics of a birth-death process based on combinatorial innovation. *J. Theor. Biol.* 491, 110187, (2020).

22. M. Sephton. Organic compounds in carbonaceous meteorites. *Natural Product Reports* 19, 292–311 (2002).

23. M. Barresi and S. Gilbert. *Developmental Biology*. Oxford University Press, Twelfth Edition, 2020.

24. G. B. West and J.H. Brown, Life's universal scaling laws. *Phys. Today* 5, 36 (2004).

25. E. Schrödinger. *What Is Life?* Cambridge University Press, 1944.

26. C. Albrecht, E. Viturro, The ABCA subfamily – gene and protein structures, functions and associated hereditary diseases. *Pfluger's Archiv – European Journal of Physiology*, 453, 581 – 589 (2006).

27. M. Eigen. *From Strange Simplicity to Familiar Complexity: A Treatise on Matter, Information, Life and Thought*. Oxford University Press, 2013.

28. N. V. Reuven, E. V. Koonin, K. E. Rudd, M.P. Deutcher, The gene for the longest known Eschereichei coli protein is a member of the helicase superfamily II. *J. Bacteriol*, 117, 5393-5400 (1995).

29. D. Raup. *Extinction: Bad Luck or Bad Genes*, Norton, 1991.

30. S. Kauffman. *At Home in the Universe*. Oxford University Press, 1995.

31. R. Hanel, S. Kauffman, S. Thurner, Towards a physics of evolution: critical diversity dynamics at the edges of collapse and bursts of diversification. *Phys. Rev.* 76, 036110, (2007).

32. S. Kauffman. On quantum gravity if non-locality is fundamental. *Entropy*, 24(4), 554 (2022). https://doi.org/10.3390/e24040554